\documentclass[numberedheadings]{aipproc}
\layoutstyle{6x9}

\begin{document}

\title{Dark Energy Present and Future}

\author{Paul H. Frampton}
{address={Department of Physics and Astronomy,
University of North Carolina, Chapel Hill, NC 27599-3255, USA}}

\begin{abstract}
By studying the present cosmological data, particularly
on CMB, SNeIA and LSS, we find that the future fate of
the universe, for simple linear models of the dark energy
equation-of-state, can vary between the extremes
of (I) a divergence of the scale factor in as little as 7 Gyr; (II)
an infinite lifetime of the universe
with dark energy dominant for all future time;
(III) a disappearing dark energy where the
universe asymptotes as $t \rightarrow \infty$
to $a(t) \sim t^{2/3}$ {\it i.e.} matter domination.
Precision cosmological data
hint that a
dark energy with equation of state $w = P/\rho < -1$
and hence dubious stability is viable.
Here we discuss for any $w$ nucleation
from $\Lambda > 0$ to $\Lambda = 0$ in a
first-order phase transition.
The critical radius is argued to be at least of galactic size
and the corresponding nucleation rate glacial,
thus underwriting the dark energy's stability and rendering
remote any microscopic effect.
\end{abstract}

\maketitle

\section{Introduction}

\bigskip
\bigskip

\noindent In this talk I will focus on two issues
concerning the enigmatic dark energy which comprises
approximately 70\% of the cosmological energy density
in the present era.

The first is concerning the future fate of the dark energy
and how it depends on the equation of state.

The second is the issue of stability of the dark energy
assuming that it can undergo a first-order phase transition
to a vacuum of truly zero energy density.

\bigskip
\bigskip

\section{Fate of Dark Energy}

\bigskip
\bigskip

The cosmic concordance of data from three disparate sources:
Cosmic Microwave Background (CMB), Large Scale Structure (LSS)
and High-Red-Shift Supernovae (SNeIA) suggests that the
present values of the dark energy and matter components,
in terms of the critical density, are approximately
$\Omega_{\rm X} \simeq 0.7$ and $\Omega_{\rm M} \simeq 0.3$.
The question to which we try to make a small contribution
in this paper is to what extent precision
cosmological data will allow
us to discriminate between possible futute fates
of the Universe?

If one makes the most conservative assumption that
$\Omega_X$ corresponds to a cosmological constant
with Equation of State given by a constant
$w = p/\rho = -1$, then the future evolution of
the universe follows from the Friedmann
equation

\begin{equation}
\left( \frac{\dot{a}}{a} \right)^2 = \frac{8 \pi G \rho}{3} +
\frac{\Lambda}{3}
\label{friedmann}
\end{equation}
in which $a(t)$ is the scale factor normalized at the present time as
$a(t_0)=1$, $\rho(t)=\rho(0)a^{-3}$ is the energy density of matter
component and $\Lambda$ is constant.  Here we assumed that the universe
is flat and neglected radiation.

In such a simple, and still viable, case the behavior
of $a(t)$ for asymptotically large $t \rightarrow \infty$
is
\begin{equation}
a(t) \sim \exp \left( \sqrt {\frac{\Lambda}{3}} t \right)
\label{desitter}
\end{equation}
so that the dark energy asymptotically dominates
and the universe is blown apart in an infinite time
$a(t) \rightarrow \infty$ as $t \rightarrow \infty$.

Even assuming that $w$ is constant, however, there is a wide range
of possible $w$: according to \cite{HM} the allowed values are $-2.68
< w < -0.78$. We do not assume this result but will arrive at
a similar allowed range; the difference
is because our priors are slightly different
(we fix the cosmological parameters
$\Omega_M = 0.3, \Omega_{\Lambda} = 0.7,
\Omega_b = 0.02,$ and $h = 0.65$
instead of allowing them to vary.)
 The asymptotic behavior in Eq.(\ref{desitter}) is
very far from established by present data.
The case $w < -1$ has the property
that boosting from the dark energy rest
frame to an inertial frame with velocity satisfying $(v/c)^2 > -1/w$
leads to a negative energy density,
but this does not
violate any law of physics\footnote{
This violates the weak energy condition\cite{hawk,Ford}.
}.
Here we shall consider
some simple models for $w$, including dependence on red-shift $w(z)$,
to illustrate how far existing data are from answering the
question of the future fate of the universe.  As we will show, for a
model in which $w$ varies linearly with red-shift, present data
are consistent with extremely different futures. For examples,
in one case the
scale factor diverges\cite{caldwell,starobinsky} in finite time
\footnote{ 
Gravitationally-bound
systems could survive longer than
$t_r$ in Eq.(\ref{tr}) but such systems would
be infinitely separated from one another.
},
in just another $7$ Gyr, while in another case the energy density of dark
energy decreases eventually faster than that of matter, i.e., the dark energy
disappears and the universe reverts to being matter-dominated with
$a(t) \sim t^{2/3}$, as $t \rightarrow \infty$.

\bigskip
\bigskip

\noindent {\it Constant Equation of State.}

\bigskip
\bigskip
Here we discuss the future fate of the universe in the case of the
constant equation of state.  If we assume, to begin, that $w$ is
constant then, keeping only the dark energy term
\begin{equation}
\left( \frac{\dot{a}}{a} \right)^2 = H_0^2 \Omega_X a^{-\beta}
\label{constantw}
\end{equation}
where $\beta = 3(1+w)$.  Most authors have discussed the case with $w
\ge -1$, however the case with $w < -1$ is also possible and
discussed phenomenologically in \cite{HM,caldwell,schulz},
and in connection with string theory in
\cite{PHF}.  If $\beta < 0$,
corresponding to $w < -1$, the solution of Eq.(\ref{constantw})
diverges at a finite time $t = t^{*}$. By integrating
\begin{equation}
\int^{\infty}_{a(t_0)} a^{\beta/2 -1} = H_0 \sqrt{\Omega_X}
\int^{t^{*}}_{t_0} dt
\label{integral}
\end{equation}
one finds that the remaining time $t_{\rm r}$ before time ends
$t_{\rm r} = (t^{*} - t_0)$ is given analytically by
\begin{equation}
t_{\rm r} = \frac{2}{3 H_0} \frac{1}{\sqrt{\Omega_X} (-w-1)}
\label{tr}
\end{equation}

In Eq.(\ref{tr}), putting in $\Omega_X = 0.7$ and
$\frac{2}{3} H_0^{-1} = 9.2$ Gyr one finds for
$w = -1.5, -2.0$ and $-2.5$, respectively
$t_{\rm r} =  22,$ 11 and 7.3 Gyr.

In such a constant $w$ scenario which is consistent
with all cosmological data, the divergence of the scale factor
will occur
in a finite time period of $7$ Gyr (or more)
from now.

With respect to the Solar System, this end of time occurs
generally after the Sun has transformed into a Red Giant,
and swallowed the Earth, as is
expected approximately $5$ Gyr in the future.

\bigskip
\bigskip

\noindent {\it Equation of State Varying Linearly with Red-Shift.}

\bigskip

\noindent{\it To view the four Figures cited, see \cite{Ta}}

\bigskip

As a more general ansatz, we consider the model for $w$ depending
linearly on red-shift:\footnote{
A model with $w$ linearly depending on red-shift is also
discussed in \cite{DW} but fitting the CMB data was not
investigated. Another parametrization of $w(Z)$
is in \cite{linder}.
}

\begin{equation}
w(Z) = w(0) + C Z \theta(\zeta - Z) + C \zeta \theta(Z -\zeta)
\label{EoS}
\end{equation}

\bigskip

\noindent where the modification is cut off arbitrarily at some $Z=\zeta >
0$.
We assume $C \leq 0$ and consider the two-dimensional
parameter space spanned by the two variables $w(0)$ and $C$.

In order to discuss constraints on our phenomenological model, we
compare its prediction with experimental data from SNeIA and CMB.
We evaluate the goodness-of-fit parameter $\chi^2$ as a function of
$w(0)$ and $C$.  For SNe1A, we used a dataset consisting of 37 SNe
from \cite{SN_Riess} with the MLCS method.  To calculate the CMB power
spectrum, we used a modified version of CMBFAST \cite{CMBFAST}. For
the analysis of CMB, we used experimental data from COBE \cite{COBE},
BOOMERanG \cite{BOOMERANG}, MAXIMA \cite{MAXIMA} and DASI \cite{DASI}.
To calculate $\chi^2$, we adopt the offset log-normal approximation
\cite{BJN} and used RADPACK package \cite{RADPACK}.  We also studied
the constraint from LSS using the 2dF data \cite{LSS}, and found that
it does not give severe constraint on the parameters $w(0)$ and $C$.

For an illustration, we take the cosmological parameters as
$\Omega_{\rm b}h^2=0.02, \Omega_{\rm M}=0.3, \Omega_{\rm X}=0.7$ and
$h=0.65$, and the initial power spectra are assumed to be scale
invariant in all numerical calculation in this paper.

To set the stage, let us first use only the SNe1A data to constrain
the parameters $w(0)$ and $C$. The result is shown for $\zeta = 2$ in
Figure 1 where the 99 \% C. L. allowed region is the region between
the two dashed lines shown. We may remark three distinct regions:

\noindent (I) $w(0) < (C - 1)$. In this case there is divergence of
the scale factor, at a finite future time.

\noindent (II) $(C - 1) \leq w(0) < C$. Here the lifetime of the universe
is infinite. The dark energy dominates
over matter, as now, at all future times.

\noindent (III) $C \leq w(0)$. The lifetime of the universe is again
infinite but after a finite time the dark energy will disappear
relative to the dark matter and matter-domination will be
re-established with $a(t) \sim t^{2/3}$.

\bigskip
\bigskip

When we add the constraints imposed by the CMB data, the allowed
region is smaller as shown in Figure 2, plotted for $\zeta =
0.5$. Such a small $\zeta$ still allows all three future possibilities
(I), (II) and (III).  For somewhat larger $\zeta$ only possibilities
(I) and (II) are allowed in this particular parameterization.

\bigskip
\bigskip

The case $\zeta = 2$ is exhibited in more detail for different values
of $w(0)$ and $C$ in Figures 3 and 4.  Figure 3 shows the variation of
the transition red-shift $Z_{\rm tr}$ where deceleration changes to
accelerated cosmic expansion defined by $q(Z_{\rm tr}) = 0$. From the
figure, we can read off that $Z_{\rm tr}$ becomes smaller as $C$
becomes more negative for fixed $w(0)$; this is because the epoch where dark
energy becomes the dominant component of the Universe becomes later.
This affects the magnitude-red shift relation of high-Z
supernovae.  In Figure 4, the magnitude-red shift relation is shown
for the SNeIA data \cite{SN_Riess} along with the prediction of our
phenomenological model for $\zeta = 2$.  The magnitudes are calculated,
as usual, relative to the empty universe Milne model with
$\Omega_M =0, ~~ \Omega_X = 0$ and $\Omega_k = 1$.  One can expect that
high-redshift SNe would appear dimmer if $C$ were more negative.

\bigskip
\bigskip

To return to our main point, let us assume that more precise cosmological
data
will allow an approximate determination of $w(Z) = f(Z)$ as a function of
$Z$
for positive $Z > 0$. Then to illustrate the possible future evolutions
write:

\begin{equation}
w(Z) = f(Z)\theta(Z) + (f(0) + \alpha Z) \theta(-Z)
\label{EoS2}
\end{equation}
In this case, the future scenarios
(I), (II) and (III) occur respectively
for $\alpha > (f(0)+ 1)$, $(f(0) + 1) > \alpha > f(0)$
and $\alpha < f(0)$.

Present data are consistent with a simple cosmological constant
$f(Z) = -1$ in Eq.(\ref{EoS2}) in which
case the divergence of the scale factor
occurs for $\alpha > 0$, the infinite-time
dark energy domination for $0 > \alpha > -1$, and
disappearing dark energy for $\alpha < -1$.

\bigskip

Since in practice $F(Z)$ for $Z \geq 0$ will never be determined
with perfect accuracy the continuation of $w(Z)$
to future $Z < 0$ will be undecidable from observation as will therefore
be the ultimate fate of the Universe.

\bigskip
\bigskip

\section{Stability Issues for Dark Energy}

\bigskip
\bigskip

\noindent The equation of state for the dark energy component
in cosmology has been the subject of much recent
discussion
\cite{PHF,Melchiorri,Schuecker,Hannestad,Caldwell,Bastero,Dicus,Ta,CHT}
Present data are consistent
with a constant  $w(Z) = -1$ corresponding to
a cosmological constant. But the data allow a present value
for
$w(Z=0)$ in the range
$- 1.38 < w(Z=0) < -0.82$ \cite{Melchiorri}.
If one assumes, more generally, that $w(Z)$ depends on $Z$
then the allowed range for $w(Z=0)$ is approximately the
same\cite{Ta}. In the present article we shall
forgo this greater generality as not relevant.
Instead, in the present article we address the question
of stability for a dark energy with constant
$w(Z) < -1$.

\bigskip
\bigskip

\noindent {\it Interpretation as a limiting velocity}

\bigskip
\bigskip

\noindent Consider making a Lorentz boost along the
1-direction with velocity $V$ (put c = 1). Then
the stress-energy tensor which in the dark energy
rest frame has the form:
\begin{equation}
T_{\mu\nu} = \Lambda
\left(
\begin{array}{cccc}
1 & 0 & 0 & 0 \\
0 & w & 0 & 0 \\
0 & 0 & w & 0 \\
0 & 0 & 0 & w
\end{array}
\right)
\label{restframe}
\end{equation}
is boosted to $T^{'}_{\mu\nu}$ given by

\begin{equation}
T^{'}_{\mu\nu} =
\Lambda
\left(
\begin{array}{cccc}
1 & V & 0 & 0 \\
V & 1 & 0 & 0 \\
0 & 0 & 1 & 0 \\
0 & 0 & 0 & 1
\end{array}
\right)
\left(
\begin{array}{cccc}
1 & 0 & 0 & 0 \\
0 & w & 0 & 0 \\
0 & 0 & w & 0 \\
0 & 0 & 0 & w
\end{array}
\right)
\left(
\begin{array}{cccc}
1 & V & 0 & 0 \\
V & 1 & 0 & 0 \\
0 & 0 & 1 & 0 \\
0 & 0 & 0 & 1
\end{array}
\right)
=
\Lambda
\left(
\begin{array}{cccc}
1+V^2w & V(1+w) & 0 & 0 \\
V(1+w) & V^2+w & 0 & 0 \\
0 & 0 & w & 0 \\
0 & 0 & 0 & w
\end{array}
\right)
\label{boost}
\end{equation}

\noindent We learn several things by studying Eq.(\ref{boost}).
First, consider the energy component $T_{00}^{'} = 1 + V^2w$.
Since $V < 1$ we see that for $w > -1$ this is positive $T_{00}^{'} > 0$
and the Weak Energy Condition (WEC) is respected\cite{hawking}.
For $w=-1$, $T_{00}^{'} \rightarrow 0$ as $V \rightarrow 1$ and
is still never negative. For $w< -1$, however, we see that
$T_{00}^{'} < 0$ if $V^2 > -(1/w)$ and this violates the WEC
and is the first sign that the case $w < -1$ must be studied
with great care.
Looking at the pressure component $T_{11}^{'}$ we see the special
role of the case $w = -1$ because
$w = T^{'}_{11}/T^{'}_{00}$ remains Lorentz invariant
as expected for a cosmological constant. Similarly
the off-diagonal components $T_{01}^{'}$
remain vanishing only in this case.
The main concern is the negativity of $T^{'}_{00} < 0$
which appears for $V^2 > -(1/w)$.
One possibility is that it is impossible
for $V^2 > -(1/w)$. The highest velocities known are
those for the highest-energy cosmic
rays which are protons with energy $\sim 10^{20}eV$.
These have $\gamma = (1 - V^2)^{-1/2} \sim 10^{11}$
corresponding to $V \sim 1 - 10^{-22}$. This would imply
that:
\begin{equation}
w > -1 - 10^{-22}
\label{sillylimit}
\end{equation}
which is one possible conclusion.

\bigskip
\bigskip

\noindent {\it First-Order Phase Transition and Nucleation Rate}.

\bigskip
\bigskip

\noindent But let us suppose, as hinted at by
\cite{Melchiorri,Schuecker,Hannestad}
that more precise cosmological data
reveals a dark matter which
violates Eq.(\ref{sillylimit}). Then, by boosting to an inertial
frame with $V^2 > -(1/w)$, one arrives at $T^{'}_{00} < 0$
and this would be a signal for vacuum instability\cite{PHF76}.
If the cosmological background
is a Friedmann-Robertson-Walker (FRW) metric the physics
is Lorentz invariant and so one should be able to see evidence
for the instability already in the preferred frame
where $T_{\mu\nu}$ is given by Eq.(\ref{restframe}).
This goes back to work in the 1960's and 1970's
where one compares the unstable vacuum to a superheated
liquid. As an example, at one atmospheric
pressure water can be heated carefully
to above $100^0$ C without boiling.
The superheated water is metastable and attempts to nucleate
bubbles containing steam. However, there is an energy
balance for a three-dimensional bubble between the positive
surface energy $\sim R^2$ and
the negative latent heat energy of the
interior $\sim R^3$ which
leads to a critical radius below which
the bubble shrinks away and above which
the bubble expands and precipitates boiling\cite{Langer1,Langer2}.
For the vacuum the first idea in \cite{PHF76}
was to treat the spacetime vacuum as a
four-dimensional material medium just like superheated water.
The second idea in the same paper
was to notice that a hyperspherical bubble expanding at the speed of light
is the same to all inertial observers. This Lorentz invariance
provided the mathematical  relationship
between the lifetime for unstable vacuum decay and
the critical radius of the four-dimensional
bubble or instanton.

In the rest frame, the energy density is
\begin{equation}
T_{00} = \Lambda \sim (10^{-3} eV)^4
\sim ({\rm 1 mm})^{-4}
\label{Lambda}
\end{equation}
since $10^{-3} eV \sim (1 {\rm mm})^{-1}$.

In order to make an estimate of the
dark energy decay lifetime
in the absence of a known potential,
we can proceed by assuming (without motivation from observation)
that
{\it there is a first-order phase transition possible between the
$\Lambda = (10^{-3} eV)^4$ ``phase'' and a $\Lambda = 0$ ``phase''}.
This hypothesized decay is the Lorentz invariant
process of a hyperspherical bubble
expanding at the speed of light, the same for all
inertial observers.
Let the radius of this hypersphere be R, its energy density be $\epsilon$
and its surface tension be $S_1$. Then according to \cite{PHF76}
the relevant instanton action is
\begin{equation}
A = -\frac{1}{2} \pi^2 R^4 \epsilon + 2 \pi^2 R^3 S_1
\label{action}
\end{equation}
where $\epsilon$ and $S_1$ are the volume and
surface energy densities, respectively.
The stationary value of this action is
\begin{equation}
A_m = \frac{27}{2} \pi^2 S_1^4 /\epsilon^3
\label{stationaryA}
\end{equation}
corresponding to the critical radius
\begin{equation}
R_m = 3S_1 / \epsilon
\label{Rcritical}
\end{equation}
We shall assume
that the wall thickness is negligible compared to
the bubble radius.
The number of vacuum nucleations in the
past lightcone is estimated
as
\begin{equation}
N = (V_u \Delta^4) exp ( - A_m)
\label{nucleations}
\end{equation}
where $V_u$ is the 4-volume of the past and
$\Delta$ is the mass scale relevant to the
problem.
This vacuum decay picture led
to the proposals of inflation\cite{Guth},
for solving the horizon, flatness and monopole problems
(only the horizon problem was generally known at
the time of \cite{PHF76}).
None of that work addressed why the true vacuum has zero energy.
Now that the observed vacuum has non-zero energy
density
$+ \epsilon \sim (10^{-3} eV)^4$
we may interpret it as a false vacuum lying
above the true vacuum with $\epsilon = 0$.
In order to use the full power
of Eq.(\ref{nucleations})
taken from \cite{PHF76,PHF77} and the
requirement $N \ll 1$ we need
to estimate the three mass-dimension parameters
$\epsilon^{1/4}, S_1^{1/3}$ and $\Delta$ therein
and so we discuss these three scales in turn.

The easiest of the three to select is $\epsilon$. If we imagine a
tunneling through a barrier between a false vacuum
with energy density $\epsilon$ to a true vacuum at energy density zero
then the energy density inside the bubble will
be $\epsilon = \Lambda = (10^{-3} eV)^4$. No other
choice is reasonable.
As for the typical mass scale $\Delta$ in the prefactor
of Eq. (\ref{nucleations}), the value of $\Delta$
does not matter very much because it appears in a power
rather than an exponential so let us put (the reader
can check that the
conclusions do not depend on this choice)
$\Delta = \epsilon^{1/4} = (1 mm)^{-1}$ whereupon the
prefactor in Eq.(\ref{nucleations}) is $\sim 10^{116}$.
The third and final scale to discuss
is the surface tension, $S_1$.
Here we appeal to comparison of spontaneous decay to stimulated
decay. The former dictates $N \ll 1$ in
Eq.(\ref{nucleations}): the latter requires further discussion.

Spontaneous dark energy decay brings us to the question of whether such
decay can be initiated in an environment existing
within our Universe. The question is analogous to
one of electroweak phase transition in
high energy particle collision.
This was first raised in \cite{PHF76}
and revisited for cosmic-ray collisions
in \cite{Hut}. That was in the context of
the standard-model Higgs vacuum and the
conclusion is that high-energy colliders
are safe at all present and planned foreseeable energies because
much more severe conditions have already
occurred (without disaster) in cosmic-ray
collisions within our galaxy.
More recently, this issue has been addressed
in connection with fears
that the Relativistic Heavy Ion Collider
(RHIC) might initiate a diastrous
transition
but according to careful analysis\cite{JBWS,AdR}
there was no such danger.

The energy density involved for dark energy is
some 58 orders of magnitude smaller [$(10^{-3}eV)^4$
compared to $(300 GeV)^4$] than for the electroweak
case and so the nucleation scales are completely different.
One is here led away from microscopic towards
astronomical size scales.

The energy density of Eq.(\ref{Lambda}) is so readily exceeded
that the critical radius cannot be
microsopic. Think first of a macroscopic scale {\it e.g. }
1 meter and consider a magnetic field
practically-attainable in bulk on Earth
such as 10 Tesla. Its energy density
is given by

\begin{equation}
\rho_{mag} = \frac{1}{\mu_0} {\cal B}^2
\label{magnet}
\end{equation}
Using the value $\mu_0 = 4 \pi \times 10^{-7} N A^{-2}$
and $1 T = 6.2 \times 10^{12} (MeV.s.m^{-2})$ leads
to an energy density
$\rho_{mag} = 2.5 \times 10^{17} eV/(mm)^3$, over 20 orders
above the value of Eq.(\ref{Lambda}) for the interior
of  nucleation. Magnetic fields in bulk exist
in galaxies with strength $\sim 1 \mu G$ and the
rescaling by ${\cal B}^2$ then would
give
$\rho_{mag} \sim (2.5 \times 10^{-5} eV)(mm)^{-3}$,
slightly below the value in Eq.(\ref{Lambda}).

{\it Assuming} the
dark energy
can exchange energy with magnetic energy density
the observed absence of stimulated decay would then
imply a critical radius of at least
galactic size, say, $\sim 10 kpc.$
Using Eq(\ref{Rcritical}) then gives for the surface tension
$S_1 > 10^{23} (mm)^{-3}$ and number
of nucleations in Eq.(\ref{nucleations})
$N < exp (- 10^{92})$. The spontaneous decay is thus glacial.
Note that the dark energy has appeared only recently
in cosmological time and has never interacted with
background radiation of comparable energy density.
Also, this nucleation argument does not require $w < -1$.

\bigskip
\bigskip

\section{Discussion}

\bigskip
\bigskip

As a first remark, since the critical radius $R_m$
for nucleation is astronomical, it appears that
the instability cannot be triggered by any
microscopic process.
While it may be comforting to know that the
dark energy is not such a doomsday phenomenon, it
also implies at the same time the dreadful conclusion
that dark energy may have no
microscopic effect.
If any such microscopic effect in a terrestrial
experiment could be
found, it would be crucial in
investigating the dark energy phenomenon.
We note that the present arguments
are less model-dependent than those in \cite{CHT}.

In closing one may speculate
how such stability arguments may evolve.
One may expect most conservatively that
the value $w = -1$ will eventually be established empirically in which case
both quintessence and the ``phantom menace''
will be irrelevant.
In that case, indeed for any $w$, we may still hope that dark energy will
provide the first connection
between string theory and the real
world as in {\it e.g.} \cite{Bastero}.
Even if precise data do establish $w < -1$,
as in the ``phantom menace'' scenario, the
dark energy stability issue is still under control.

\bigskip
\bigskip

\section*{Acknowledgement}

\bigskip
\bigskip

I thank C.N. Leung, S. Tovey and R. Volkas
for organizing this stimulating workshop.
This work was supported in part
by the Department of Energy
under Grant Number
DE-FG02-97ER-410236.

\end{document}